\documentclass[12pt,letterpaper]{IEEEtran}
\usepackage[pdftex]{graphicx}
\pdfoutput=1
\usepackage{amssymb,latexsym}
\usepackage{amsmath}
\usepackage{graphicx}
\usepackage[justification=centering]{caption}
\usepackage[style=ieee]{biblatex}
\usepackage[labelsep=period]{caption}
\usepackage[margin=1in]{geometry}
\AtEveryBibitem{%
	\clearfield{issn} % Remove issn
	\clearfield{isbn} % Remove issn
	\clearfield{doi} % Remove doi
	\clearfield{urldate}
	\clearfield{urlyear}
	\clearfield{urlmonth}
	
	\ifentrytype{online}{}{% Remove url except for @online
		\clearfield{url}
		\clearfield{note}
		\clearfield{url}
	}
}
{\footnotesize
	\bibliography{bibfile}}
\usepackage{breqn}
\usepackage{amssymb}
\usepackage{algorithm,algorithmic}
\usepackage{color}
\usepackage{lipsum,graphicx,subcaption}
\usepackage{float}
\ifCLASSINFOpdf
\else
\fi
\onecolumn

\begin{document}
	\title{LED Selection and MAP Detection for Generalized LED Index Modulation }
\author{Manh~Le~Tran, Sunghwan~Kim, Thomas~Ketseoglou, \IEEEmembership{Senior Member, IEEE} and Ender~Ayanoglu, \IEEEmembership{Fellow, IEEE}% <-this % stops a space
	\thanks{Le~Tran~Manh, Sunghwan~Kim are with  the  School  of  EE,  University  of  Ulsan,  93, Daehak-ro, Nam-gu, Ulsan, 44610, Korea (e-mail: tranmanh@mail.ulsan.ac.kr; sungkim@ulsan.ac.kr)}% <-this % stops a space
	\thanks{T. Ketseoglou is with the Electrical and Computer Engineering Department, California State Polytechnic University, Pomona, CA 91768 USA (e-mail: tketseoglou@cpp.edu)}% <-this % stops a space
	\thanks{E. Ayanoglu is with the Center for Pervasive Communications and Computing, Department of Electrical Engineering and Computer Science, University of California at Irvine, Irvine, CA 92697 USA (e-mail:ayanoglu@uci.edu)}% <-this % stops a space
	
	\markboth{Journal of ...}%
	{Shell \MakeLowercase{\textit{et al.}}: Bare Demo of IEEEtran.cls for IEEE Journals}}	
\maketitle	
\begin{abstract}
In this paper, we propose light-emitting diode (LED) selection that can be applied not only to the conventional Multiple-Input Multiple-Output (MIMO) case, but also to a larger MIMO configuration of generalized LED index modulation (GLIM) system with optical orthogonal frequency division multiplexing (OFDM) in visible light communication (VLC). Moreover, we derive a simplified implementation of the maximum a posteriori (MAP) detector when the number of LEDs is an even number larger than four. Simulation results show that the performance of MAP and LED selection is better than other detection algorithms for larger even numbers of LEDs and conventional GLIM for $4\times4$ transmission, respectively.
\end{abstract}
\IEEEpeerreviewmaketitle
\section{Introduction}
\IEEEPARstart{V}{isible} light communication (VLC) which utilizes optical wireless communication technology via high-power and high-speed modulated light-emitting diodes (LEDs) as the transmitters and photo detectors (PDs) as the receivers has the advantages of a high transmission rate, energy preservation, and secure data communication \cite{komine_fundamental_2004}. As an emerging technology, there have been many studies on multiple-input multiple-output VLC (MIMO-VLC) systems, including both theoretical analysis and experimental demonstrations \cite{zeng_high_2009}, \cite{fath_performance_2013}.

MIMO-based VLC schemes significantly suffer from performance degradation because indoor optical wireless channels have a strong correlation \cite{fath_performance_2013}. There are efforts to decorrelate the channels include altering the angle of PDs \cite{fahamuel_improved_2014}, \cite{nuwanpriya_indoor_2015}, \cite{he_performance_2015}, using imaging lens and a detector array \cite{zeng_high_2009}, or power allocation between transmitters \cite{fath_performance_2013}.

Lately, orthogonal frequency division multiplexing (OFDM) with its high spectral efficiency and robustness to channel impairments has been frequently used in optical wireless communications \cite{armstrong_ofdm_2009} such as DC-biased optical OFDM (DCO-OFDM) \cite{tan_near-optimal_2016} and asymmetrically clipped optical OFDM (ACO-OFDM) \cite{armstrong_power_2006}. These schemes also have trade-offs between various parameters that are related to system performance such as power efficiency, data rate, number of the OFDM sub-carriers.

Furthermore, optical spatial modulation (OSM) in VLC exploits the spatial domain as a solution to reduce interferences between transmitters \cite{mesleh_optical_2011}. Non-DC-biased OFDM (NDC-OFDM) in \cite{li_non-dc-biased_2013} was proposed to transmit positive and negative signals separately without DC bias while non-Hermitian symmetry OFDM (NHS-OFDM) in \cite{chen_non-hermitian_2017} required DC bias to transmit real and image parts of signals by two LEDs. Recently, the GLIM and extended GLIM techniques for $4 \times 4$ MIMO-based OFDM systems were proposed in \cite{basar_generalized_2016}, \cite{yesilkaya_optical_2017} to overcome the need for Hermitian symmetry or DC bias, which  are generally demanded for OFDM-based VLC systems. Performance of the $4\times4$ case was presented in \cite{yesilkaya_optical_2017}. However, a generalization of the results in \cite{yesilkaya_optical_2017} to a larger number of LEDs was not discussed and application of the GLIM in \cite{yesilkaya_optical_2017} to larger number of LEDs system showed worse performance.

In this letter, we expand GLIM to even numbers of LEDs larger than four, and derive a much simpler MAP-based demodulation algorithm at the receiver. We also propose LED selection to decorrelate MIMO channels for performance enhancement. The LED selection technique can be applied to the $4\times 4$ GLIM case with various channels for channel decorrelation, and show better decorrelation performance when the number of LEDs is larger than four. Simulation results show that when the number of LEDs increases, the proposed system achieves much better performances than the original GLIM system.

~~

~~
\section{Proposed GLIM System}
We introduce the system model of the proposed GLIM scheme based on the original GLIM system \cite{yesilkaya_optical_2017}, which focuses on MIMO transmission $n_{\text T} \times n_{\text R}$ where $n_{\text T}$, $n_{\text R}$ are number of LEDs and PDs, respectively. The general explanation of LED selection for large numbers of LEDs will be discussed in the next subsection. 
\begin{figure}[h] 
	\centering
	\includegraphics[width=4in]{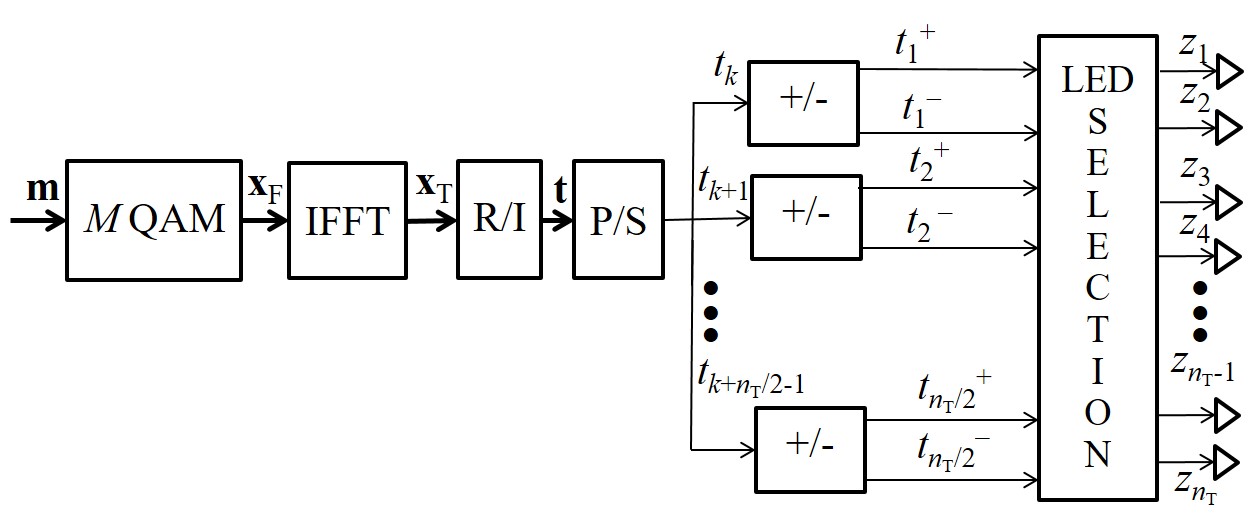}
	\caption{Proposed transmitter model}
	\label{Fig:LED_4x4}
\end{figure}
\subsection{System Model}
The proposed system model has multiple LEDs and PDs. Let $n_{\text T}$ and $n_{\text R}$ be the number of LEDs and PDs, respectively, where $n_{\text T}$ is assumed to be an even number for GLIM.
Let the vector $\bold z$ be the transmitted signal from the $n_{\text T}$ LEDs at time $k$, written as $\bold z = [z_{1} \cdots {z_{n_{\text T}}}]^T$.
The time-domain OFDM signals $x_l$ in ${{\bf{x}}_{\bf{T}}}$, $l$=$1,\cdots,N$ are separated into their real and imaginary parts as $x_l =x_{l, {R}}+jx_{l,{I}}$ and converted into a time domain signal vector as ${\bf{t}} = \left[ {{x_{1,R}},{x_{1,I}},{x_{2,R}},{x_{2,I}},...,{x_{N,R}},{x_{N,I}}} \right]$. Then, the `P/S' box chooses $n_{\text T}/2$ elements in ${\bf{t}}$ in consecutive order as $\left[ {{t_k},{t_{k + 1}},...,{t_{k + {n_{\text{T}}}/2 - 1}}} \right]$. The index $k$ is in the set $\{1,{n_{\text{T}}}/2 + 1,{n_{\text{T}}} + 1,...\}$. Let sgn(a) be
\[
\textrm{sgn(a)}=\left\{ \begin{matrix} 1, & \text{if~} a \geq 0 \\ -1, &\text{if~} a<0. \end{matrix}  \right.
\]
The real signal $t_{k+l}$ after `+/-' box can be transmitted as
\[
t_{l}^+=\frac{\textrm{sgn}(t_{k+l-1})+1}{2} t_{k+l-1},
t_{l}^-=\frac{\textrm{sgn}(t_{k+l-1})-1}{2} t_{k+l-1}. \]

The signals $t_{l}^+$ and $t_{l}^-$ with $l$=$1, 2,\cdots,n_{\text T}/2$ can be mapped to $\bold z$. Let the LED set $L$ be $\{1, 2, \cdots, n_{\text T} \}$. Then $n_{\text T}/2$ signals can be transmitted through ${n_{\text T}/2}$ sets of ${\bold s}_1, \cdots, {\bold s}_{{n_{\text T}}/2}$, which are subsets of $L$. Moreover, the sets ${\bold s}_1, \cdots, {\bold s}_{{n_{\text T}}/2}$ are mutually exclusive and collectively exhaustive. Each of the vectors ${\bold s}_1, \cdots, {\bold s}_{n_{\text T}/2} $ will represent the signals $\{t_{l}^+$, $t_{l}^-\}$, $l$=$1,2,\cdots,n_{\text T}/2$.
For example, if ${\bold s}_1 =\{1, 3\}$, ${\bold s}_2 =\{2, 4\}$, ${\bold s}_3=\{5, 7\}$ and ${\bold s}_4 =\{6, 8\}$, then ${z_{1}} = t_{1}^ + $, ${z_{2}} = t_{1}^ - $, ${z_{3}} = t_{2}^ + $, ${z_{4}} = t_{2}^ - $, ${z_{5}} = t_{3}^ + $, ${z_{6}} = t_{3}^ - $, ${z_{7}} = t_{4}^ + $ and ${z_{8}} = t_{4}^ - $.

The signal $\bold z$ is transmitted over the $n_{\text T} \times n_{\text R}$ optical MIMO channel as $\bold{y}=\bold{H} \bold z +\bold n$, where $\bold{y}= [ y_{1}, \cdots,y_{n_{\text R}}]$, is the received signal. The signal $\bold z$ contains the electrical signals obtained from PDs, and the real-valued additive white Gaussian noise  $\bold n$ whose elements are distributed as $\mathcal{N}(0,\,\sigma_w^{2})$.
\subsection{LED Selection in GLIM}
The LED selection technique is used to improve the performance of the GLIM system with the MAP detector which is affected by the correlation between LEDs. In particular, a candidate ${{\bf{C}}_i}$ is a map of the signals to LEDs. In the $4 \times 4$ case \cite{yesilkaya_optical_2017}, ${{\bf{C}}_i} = \left\{ {{{\bf{s}}_1},{{\bf{s}}_2}} \right\}$ with ${\bold s}_1 =\{1, 2\}$ and ${\bold s}_2 =\{3, 4\}$. Another candidate for $4 \times 4$ LED selection is ${\bold s}_1 =\{1, 3\}$ and ${\bold s}_2 =\{2, 4\}$ shown in Fig.~\ref{Fig:LED_4x4}. Let ${{\bf{H}}_i}$ be an $n_{\text R} \times n_{\text T}$ channel matrix corresponding to ${{\bf{C}}_i}$. Then active channel matrix ${\bf{H}}_j^i$ with size $n_{\text R} \times n_{\text T}/2$ is a sub-matrix of ${{\bf{H}}_i}$ for $j = 1,...\,{2^{{n_{\text{T}}}/2}}$, which depends on active LEDs to transmit the signal.

When transmitted through a number of LEDs, the performance of the system heavily depends on the SNR and correlation of the channel, as this determines the condition of the MIMO channel. For convenience, the first signal $t_1^ + $ is allocated to the first LED. Then, the first element satisfies ${{\bf{s}}_1}(1) = 1$. Moreover, because of the symmetric position of the LEDs, the number of selections can be reduced.
\begin{figure}[h] 
	\centering
	~~\includegraphics[width=1.0in]{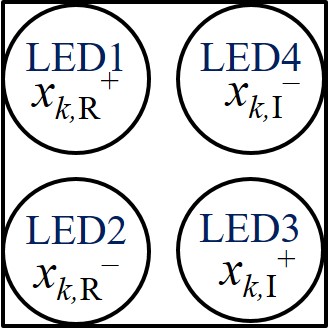}
	~~~~~~~~~~~~~~\includegraphics[width=1.0in]{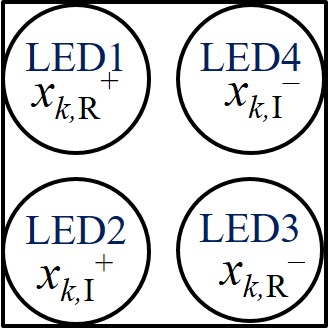}\\ ~(a) ${{\bf{C}}_1}$ with~~~~~~~~~~~~~~~~~ (b) ${{\bf{C}}_2}$ with\\~~${\bold s}_1 =\{1, 2\}, {\bold s}_2 =\{3, 4\}$~~~~~~${\bold s}_1 =\{1, 3\}, {\bold s}_2 =\{2, 4\}$
	\caption{LED selection in $4 \times 4$ transmission}
	\label{Fig:LED_4x4}
\end{figure}
The correlation of the channel as the selection of one LED affects the receiving performance of GLIM according to two factors:
a) The correlation between a couple of active-inactive LEDs channel couples ${{\bf{h}}_a}$, ${{\bf{h}}_b}$ measured by the cosine between them \cite{jiang_angles_1996} defined as
\begin{equation}
\textrm{cos}({{\bf{h}}_a},{{\bf{h}}_b}) = \frac{{{\bf{h}}_a^H{{\bf{h}}_b}}}{{\left\| {{{\bf{h}}_a}} \right\|\left\| {{{\bf{h}}_b}} \right\|}},
\end{equation}
where ${\left\| . \right\|}$ denotes the $l_2{\text{ - norm}}$ and 
b) the condition number between active channel columns $c({{\bf{H}}^i_j})$ \cite{tse_fundamentals_2005}, defined as
\begin{equation}
c({{\bf{H}}^i_j}) = \frac{{{\lambda _{\max }}}}{{{\lambda _{\min }}}},
\end{equation}
where ${\lambda _{\max }},{\lambda _{\min }}$ are the maximum and minimum singular values of the channel matrix, respectively. An algorithm to jointly optimize two terms will first remove candidates mostly impacted by an incorrect active decision, then select the candidate that can give the best channel capacity. These steps are described in Algorithm \ref{algo2}.
\begin{algorithm}[]
	\caption{Algorithm for $n_{\text T} \times  n_{\text R}$ LED selection}
	\label{algo2}
	\begin{algorithmic}[1]
		\renewcommand{\algorithmicrequire}{\textbf{Input:}}
		\renewcommand{\algorithmicensure}{\textbf{Output:}}
		\REQUIRE Channel matrix {\bfseries{H}}
		\ENSURE  Optimal channel ${{\bf{H}}^{opt}}$
		\STATE generate set ${\bf{\Theta }} = \left\{ {{\theta _1},{\theta _2}...} \right\}$ which includes all LED pairs ${\theta _1} = \left\{ {1,2} \right\},{\theta _2} = \left\{ {1,3} \right\},{\theta _3} = \left\{ {1,4} \right\},...$
		\STATE calculate ${t_i} = \textrm{cos} ({\theta _i})$ with ${\theta _i} \in {\bf{\Theta }}$
		\STATE set $\omega  = \max ({t_i})$
		\STATE if ${t_i} = \omega $ then ${\bf{\Theta }} = {\bf{\Theta }} - \{{\theta _i}\}$
		\STATE from ${\bf{\Theta }}$ generate ${\bf{U}} = \left\{ {{{\bf{C}}_1},{{\bf{C}}_2}...} \right\}$ of all candidate ${{\bf{C}}_i}$
		\FOR {${{\bf{C}}_i} \in {\bf{U}}$}
		\STATE generate active set ${V_i} = \left\{ {{\bf{H}}_1^i,...{\bf{H}}_{{2^{{n_{\text{T}}}/2}}}^i} \right\}$
		\STATE set ${\mu _i} = \mathop {\max }\limits_{{\bf{H}}_j^i \in {{\bf{V}}_i}} c({\bf{H}}_j^i)$
		\ENDFOR	
		\STATE set ${{\bf{C}}_{selection}} = \mathop {\arg \min }\limits_{{{\bf{C}}_i} \in {\bf{U}}} ({\mu _i})$
		\RETURN ${{\bf{H}}^{opt}}$ corresponding to ${{\bf{C}}_{selection}}$
	\end{algorithmic} 
\end{algorithm}
\subsection{MAP Detection}
Among the three detection algorithms, e.g., zero-forcing (ZF), minimum mean-square error (MMSE) and MAP, MAP is considered here, since its performance provides the best performance improvement \cite{yesilkaya_optical_2017}. Meanwhile, the MAP detection in \cite{yesilkaya_optical_2017} is focused on the $4\times4$ case and requires a complex mathematical representation with a higher dimensional system such as 8, 16, or more LEDs.

To employ MAP for more complex systems, we derive a much simpler representation of the solution that can be used in any GLIM system. For generality, take a VLC system with an $n_{\text T} \times n_{\text R}$ channel matrix ${\bf{H}} = \left[ {{{\bf{h}}_1}\,{{\bf{h}}_2}\,...\,{{\bf{h}}_{{n_{\text{T}}}}}} \right]$, where $l = 1,...,{n_{\text{T}}}$ is the $l{\text{-th}}$ column of ${\bf{H}}$, the received signal is 
\begin{equation}
\begin{aligned}
{\bf{y}} &= {\bf{Hz}} + {\bf{n}} = \sum\nolimits_{l = 1}^{{n_{\text{T}}}/2} \left( {{{\bf{h}}_{{{\bf{s}}_{l}}(1)}}}t_l^ +  + {{{\bf{h}}_{{{\bf{s}}_{l}}(2)}}}t_l^ -  \right)  + {\bf{n}}\\
&= \sum\nolimits_{l = 1}^{{n_{\text{T}}}/2} {{\bf{h}}_l^\Phi } {\bar t_l} + {\bf{n}},
\end{aligned}
\end{equation}
where ${\bar t_{l}} = \left| {{t_{l}}} \right|$ and ${{\bf{h}}_l^\Phi }$ is either ${{{\bf{h}}_{{{\bf{s}}_{l}}(1)}}}$ or ${{{\bf{h}}_{{{\bf{s}}_{l}}(2)}}}$, $l = 1,...{n_{\text T}}/2$. For a given set $\Phi =\left\{ {{{\bf{h}}_{1}^\Phi},...{\bf{h}}_{n_{\text T}/2}^\Phi} \right\}$, by taking into account the clipped Gaussian distribution of the transmitted signal as the prior information \cite{yesilkaya_optical_2017}, the conditional MAP estimation of ${\bf{\hat t}}$ can be obtained in a fashion similar to \cite{yesilkaya_optical_2017}, as
\begin{equation}
\begin{aligned}
{{\bf{\hat t}}} &= \arg \mathop {\min }\limits_{{\bf{\bar t}_\Phi}} {M^{MAP}}({{\bf{h}}_{{\Phi _1}}},...{{\bf{h}}_{{\Phi _{{n_{\text{T}}}/2}}}},{\bar t_1},...{\bar t_{{n_{\text{T}}}/2}},)\\
&= \arg \mathop {\min }\limits_{{{{\bf{\bar t}}}_{\bf{\Phi }}}} \big({\left\| {{\bf{y}} - {{{\bf{\bar H}}}_\Phi }{{{\bf{\bar t}}}_{\bf{\Phi }}}} \right\|^2} + 2\sigma _w^2{\left\| {{{{\bf{\bar t}}}_{\bf{\Phi }}}} \right\|^2}\big),
\end{aligned}\\
\end{equation}
where ${{\bf{\bar t}_\Phi}} = {\left[ {\bar t_1^\Phi ...\bar t_{{n_{\text{T}}}/2}^\Phi } \right]^T}$, ${{\bf{\bar H}}_\Phi } = {\left[ {{{\bf{h}}_{{\Phi _1}}}...{{\bf{h}}_{{\Phi _{{n_{\text{T}}}/2}}}}} \right]}$ is one of ${2^{{n_{\text{T}}}/2}}$ matrix candidates of size ${n_{\text R}} \times {n_{\text T}/2}$ and corresponding to the set $\Phi$.
Differentiating second part of (4) with respect to ${{\bf{\bar t}_\Phi}}$ and equating it to zero, the solution for ${{\bf{\bar t}_\Phi}}$ is
\begin{equation}
{\bf{\bar t}_\Phi} = {{\bf{A}}_\Phi }{\bf{y}},
\label{equa:find t}
\end{equation}
where ${{\bf{A}}_\Phi } = {\left( {{\bf{\bar H}}_\Phi ^T{{{\bf{\bar H}}}_\Phi } + 2\sigma _w^2{\bf{I}}} \right)^{ - 1}}{\bf{\bar H}}_\Phi ^T$. Let vector ${\left[ {\bf{a}} \right]^ + }$ denote that each element in ${\left[ {\bf{a}} \right]^ + }$ calculated as $\max \left( {0,{a_i}} \right)$ where $a_i$ is the $i{\text{-th}}$ element in $\bf{a}$. Then, due to positive value condition of light intensity, we need to take ${{\bf{\tilde t}_\Phi}} = {\left[ {{{{\bf{\bar t}_\Phi}}}} \right]^ + }$. After calculation of the MAP estimates for all cases of $\Phi$ and considering all possible active LED scenarios, the conditional MAP estimator of the GLIM-OFDM scheme decides on the most likely active LEDs by calculating the MAP estimation metric given for all scenarios. The unconditional estimates of ${{\tilde t}_{l}}$, $l = 1,...{n_{\text T}}/2$ is 
\begin{equation}
({\bf{\hat t}},\Phi)  = \arg \mathop {\min }\limits_{{{{\bf{\tilde t}}}_{\bf{\Phi }}}} {M^{MAP}}({{\bf{h}}_{{\Phi _1}}},...{{\bf{h}}_{{\Phi _{{n_{\text{T}}}/2}}}},{\tilde t_1},...{\tilde t_{{n_{\text{T}}}/2}}),
\end{equation}
where ${\hat t_l} = \tilde t_l^{\hat{\Phi}}$. By precalculating the values of ${{\bf{A}}_\Phi}$ for all cases of the set $\Phi$ beforehand and storing them at the receiver, the demodulation procedure is presented in Algorithm \ref{algo1}. 
\begin{algorithm}[H]
	\caption{MAP estimation for $n_{\text T} \times  n_{\text R}$ GLIM}
	\label{algo1}
	\begin{algorithmic}[1]
		\renewcommand{\algorithmicrequire}{\textbf{Input:}}
		\renewcommand{\algorithmicensure}{\textbf{Output:}}
		\REQUIRE Channel matrix {\bfseries{H}}
		\ENSURE  Estimated signal ${\hat{\bf{t}}}$
		\\ \textit{Initialization} :
		\STATE Generate all matrices ${{\bf{H}}_\Phi } = {\left[ {{{\bf{h}}_{{\Phi _1}}}...{{\bf{h}}_{{\Phi _{{n_{\text{T}}}/2}}}}} \right]}$ of size $n_{\text R}\times n_{\text T}/2$ corresponding to the set $\Phi$ by selecting $n_{\text T}/2$ columns of channel matrix {\bfseries{H}}\\
		Calculate matrices ${\bf{A}_\Phi}$ for all cases of $\Phi$
		\\ \textit{LOOP Process}
		\FOR {${{\bf{H}}_\Phi}$}
		\STATE Calculate ${{{\bf{\tilde t}}}_{\bf{\Phi }}} = {\left[ {{{\bf{A}}_\Phi }{\bf{y}}} \right]^ + }$ for $\Phi$ from equation (5)
		\STATE Calculate $\bf{\hat t}$, ${{\bf{H}}^{active}}$ from equation (6)
		\ENDFOR	
		\RETURN ${\hat{\bf{t}}}$ 
	\end{algorithmic} 
\end{algorithm}
\section{Numerical results}
\begin{figure}[!h] 
	\centering
	~~\includegraphics[width=2.5in]{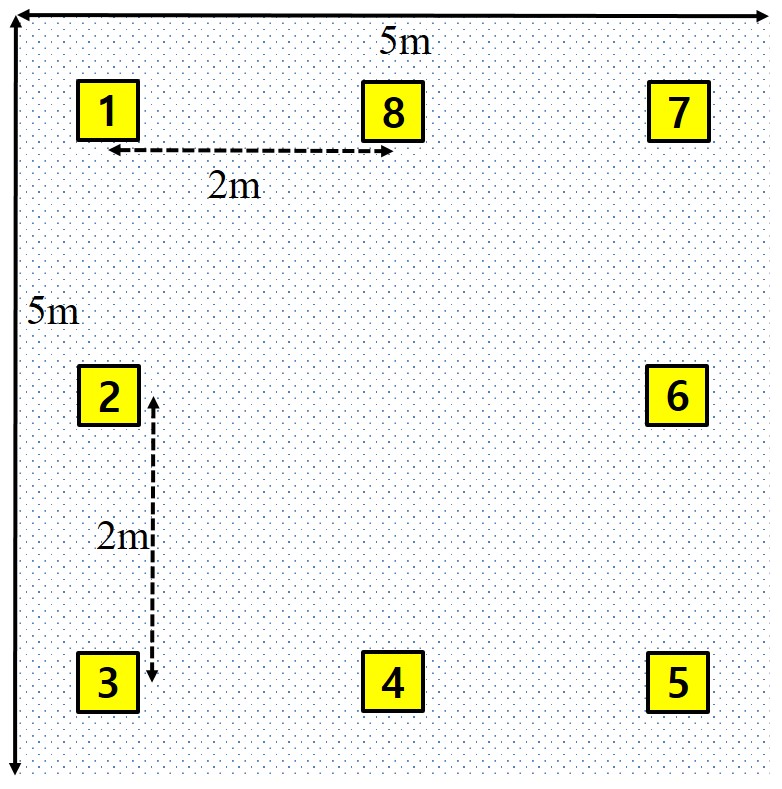}
	~\includegraphics[width=1.8in]{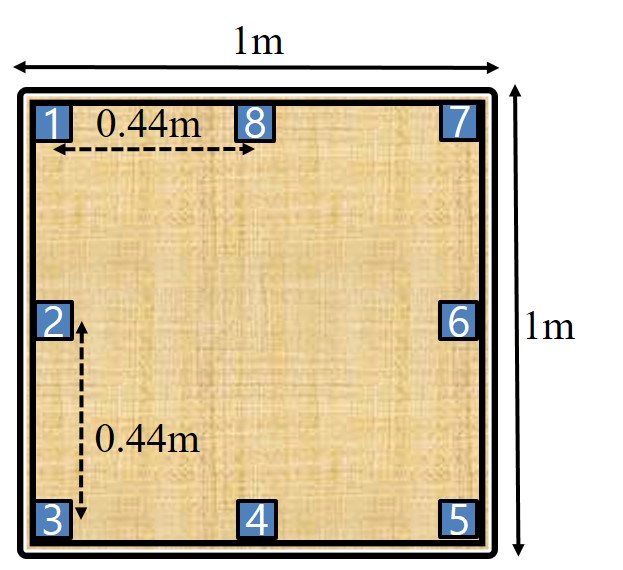}\\
	~~(a) LED position~~~~~~~(b) PD position
	\caption{Position for $8 \times 8$ GLIM scenario}
	\label{Fig:LED_8x8}
\end{figure}

We consider an $8 \times  8$ GLIM system arranged in a square shape as in Fig.~\ref{Fig:LED_8x8}, where one side of the square of the LEDs is 4 meters and the corresponding side for PDs is 1 meter. 
The transmitter uses OFDM with symbols from the 4QAM, 8QAM, or 16QAM mappings, and the receiver uses the MAP, ZF \cite{yesilkaya_optical_2017}, or MMSE \cite{tse_fundamentals_2005}.
\begin{figure}[H] 
	\centering
	\includegraphics[width=3.3in]{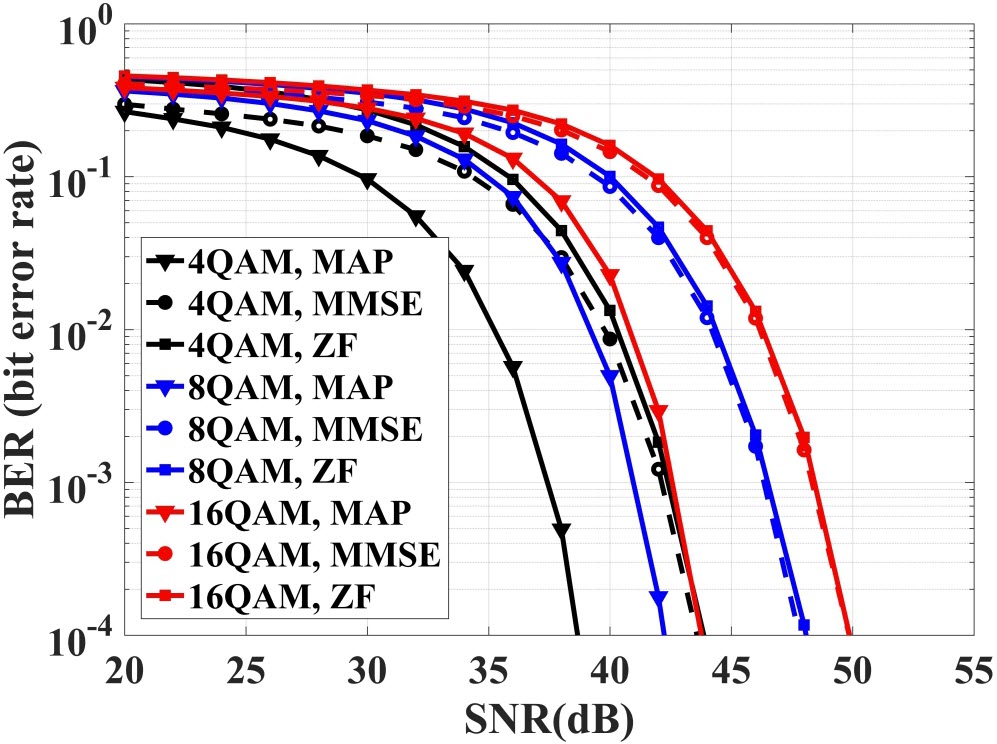}
	\caption{Performance comparison of the $8\times 8$ GLIM with MAP, ZF and MMSE demodulators}
	\label{Fig:8x8_compare}
\end{figure}
Fig.~\ref{Fig:8x8_compare} shows that the bit error rate (BER) performance of the ZF and MMSE detectors which are the same for high SNR. Meanwhile, the MAP detector still gives the best performance compared with the others, as in the $4 \times  4$ case \cite{yesilkaya_optical_2017}.
\begin{figure}[H] 
	\centering
	\includegraphics[width=3.3in]{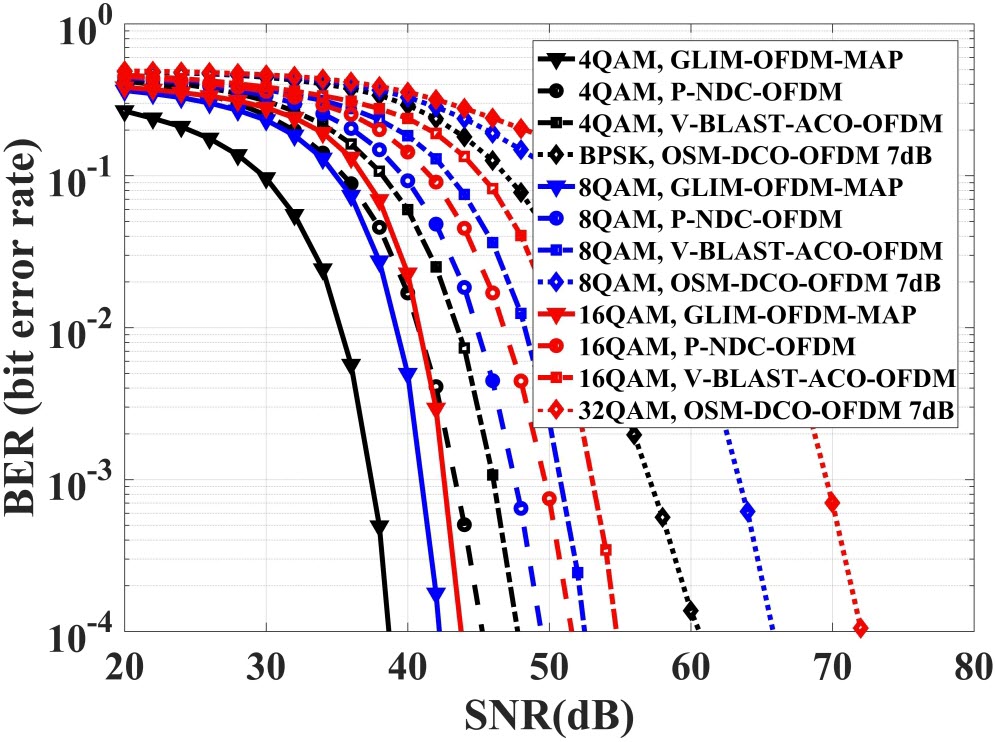}
	\caption{Performance comparison of the $8\times 8$ GLIM with other modulation schemes}
	\label{Fig:8x8_ref}
\end{figure}

In Fig.~\ref{Fig:8x8_ref} we compare the performance of GLIM with three reference systems for $8 \times 8$ MIMO-OFDM VLC transmission. To ensure fairness, reference schemes transmit same number of bits/s/Hz. 'P-NDC-OFDM', 'V-BLAST-ACO-OFDM', and 'OSM-DCO-OFDM' in Fig.~\ref{Fig:8x8_ref} stand for four parallel NDC-OFDM systems \cite{li_non-dc-biased_2013}, eight parallel vertical Bell Laboratories Layered Space-Time ACO-OFDM systems \cite{armstrong_power_2006}, and combination of DCO-OFDM \cite{tan_near-optimal_2016} with OSM \cite{mesleh_optical_2011}, respectively. BERs of GLIM with MAP are still better than $8 \times 8$ reference systems, which is similar to the $4 \times 4$ case in \cite{yesilkaya_optical_2017}.
\begin{figure*}[ht!]
	\centering
	{%
		\includegraphics[width=.31\linewidth]{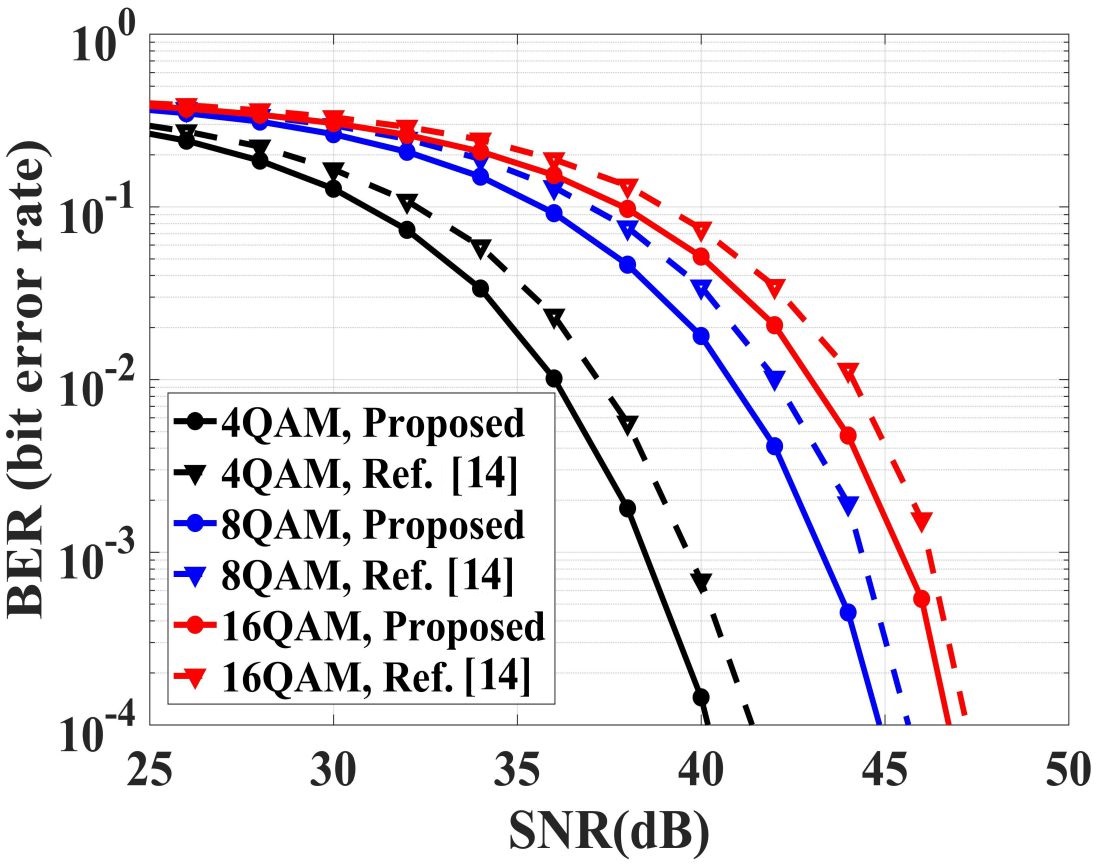}}\quad
	{%
		\includegraphics[width=.31\linewidth]{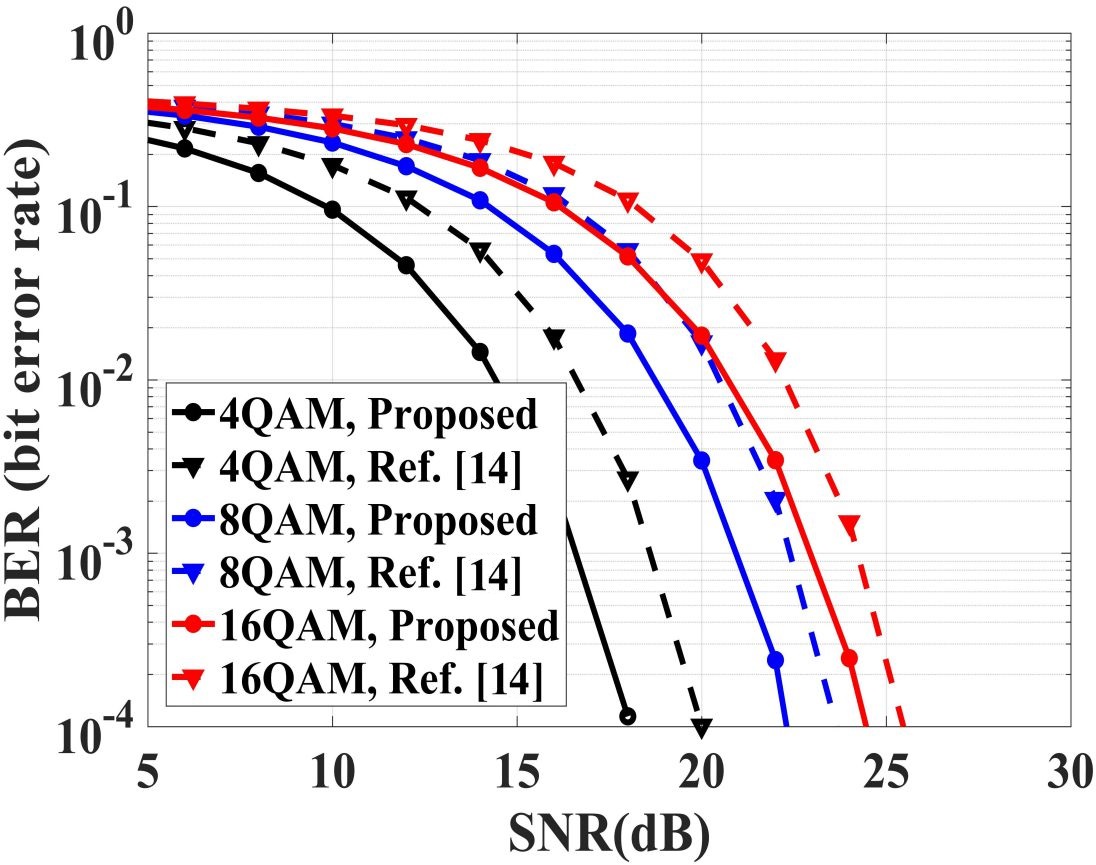}}\quad
	{%
		\includegraphics[width=.31\linewidth]{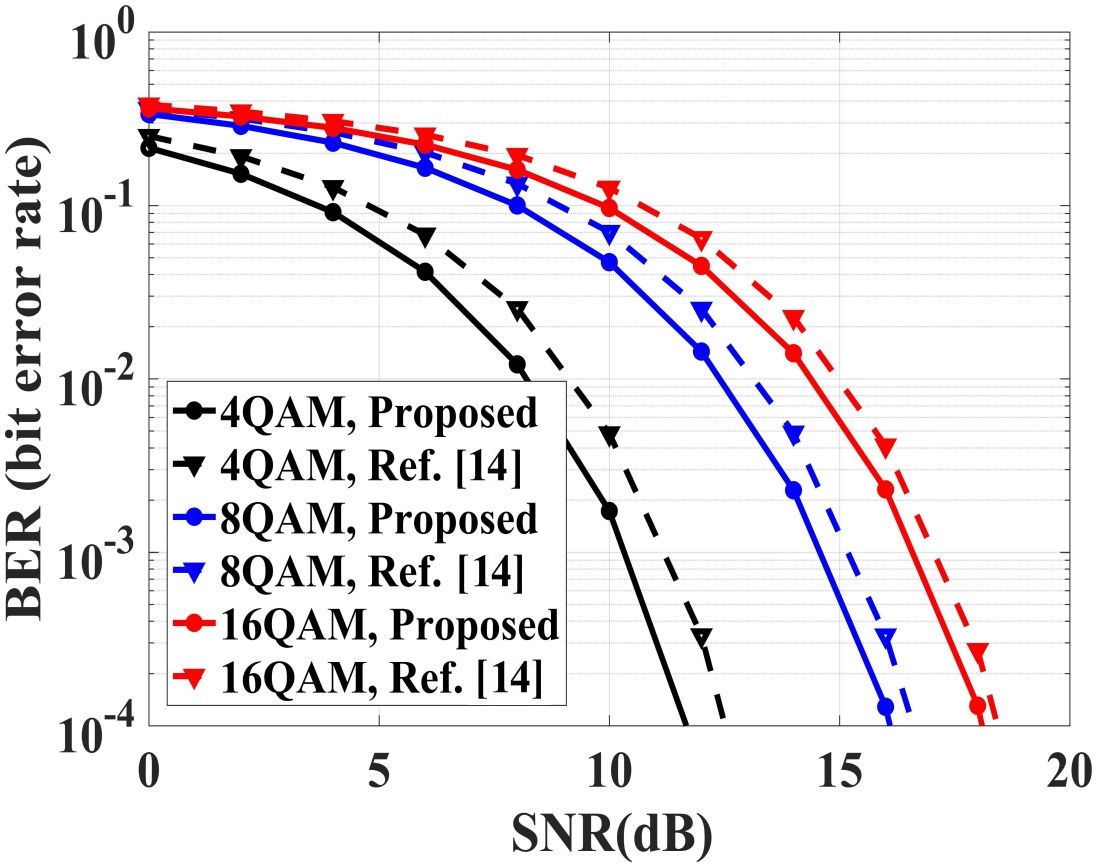}}\\
	~~(a) Physical channel~A~~~~~~~~~~~~~~~~(b) Physical channel~B~~~~~~~~~~~~~~(c)  Physical channel~C
	\caption{LED selection for the $4 \times 4$ GLIM system}
	\label{fig:4x4_selection}
\end{figure*}

To examine the effectiveness of LED selection, results of reference GLIM \cite{yesilkaya_optical_2017} and the proposed GLIM for $4 \times 4$ transmission are shown in Fig.~\ref{fig:4x4_selection} where physical channel A, physical channel B, and physical channel C are same with the ones in \cite{yesilkaya_optical_2017}. From Algorithm 1, LED selection is determined as ${\bold s}_1 =\{1, 3\}$ and ${\bold s}_2 =\{2, 4\}$. Over the three channels, BER performances of the proposed system are better than the ones in \cite{yesilkaya_optical_2017}.\\
\begin{figure}[h] 
	\centering
	\includegraphics[width=3.3in]{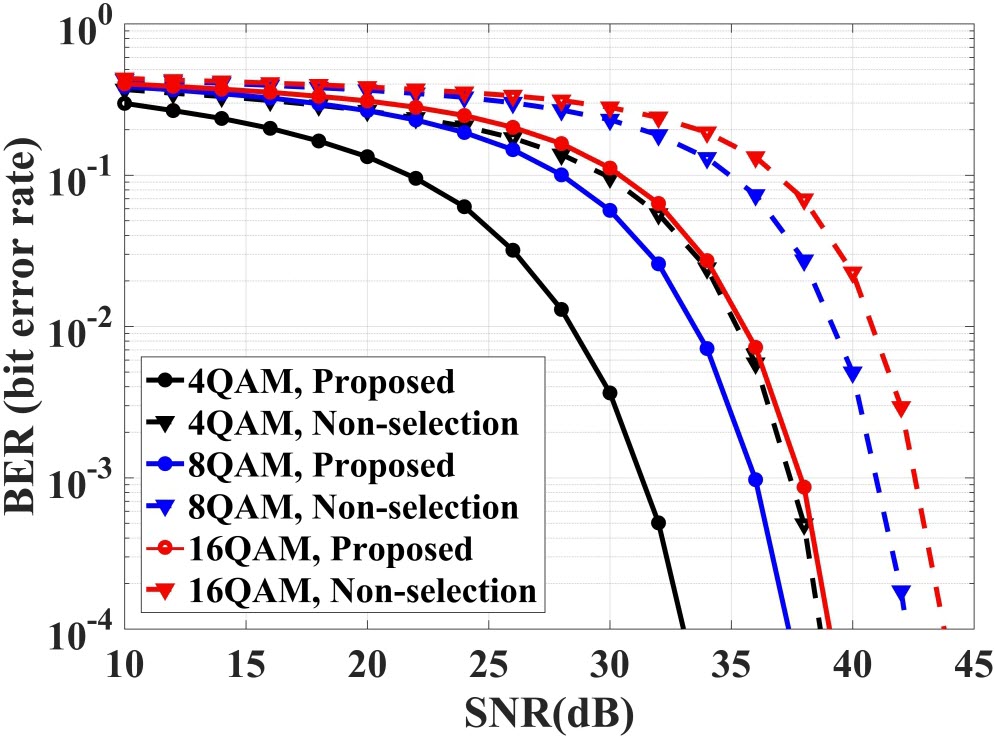}
	\caption{LED selection for the $8 \times 8$ GLIM system}
	\label{Fig:8x8_selection}
\end{figure}
To demonstrate LED selection can be implemented with larger number of LED systems, performances of GLIM with and without selection for $8 \times 8$ transmission are shown in Fig.~\ref{Fig:8x8_selection}. No LED selection means ${\bold s}_1 =\{1, 2\}$, ${\bold s}_2 =\{3, 4\}$, ${\bold s}_3 =\{5, 6\}$ and ${\bold s}_4 =\{7, 8\}$. From Algorithm 1, the LED selection is ${\bold s}_1 =\{1, 3\}$, ${\bold s}_2 =\{2, 4\}$, ${\bold s}_3 =\{5, 7\}$ and ${\bold s}_4 =\{6, 8\}$. Performances of the proposed GLIM are about 5dB better than ones without LED selection for three QAM modulations.
\section{Conclusion}
Simplified MAP and LED selection are proposed to enhance the transmission when the number of LEDs is even and larger than four. These schemes are not just extensions of $4 \times 4$ transmission, but they are also more efficient techniques for high order transmission. Simulation results show that the proposed MAP and LED selection lead to significant performance improvements in VLC systems.
\ifCLASSOPTIONcaptionsoff
\newpage
\fi
\renewcommand*{\bibfont}{\footnotesize}
\printbibliography
\end{document}